# Cortical Factor Feedback Model for Cellular Locomotion and Cytofission

Shin I. Nishimura[1]*, Masahiro Ueda[2,3], Masaki Sasai[1]

1 Department of Computational Science and Engineering, Nagoya University, Nagoya, Japan, 2 Laboratories for Nanobiology, Graduate School of Frontier Biosciences, Osaka University, Suita, Osaka, Japan, 3 JST, CREST, Suita, Osaka, Japan

**Abstract**

Eukaryotic cells can move spontaneously without being guided by external cues. For such spontaneous movements, a variety of different modes have been observed, including the amoeboid-like locomotion with protrusion of multiple pseudopods, the keratocyte-like locomotion with a widely spread lamellipodium, cell division with two daughter cells crawling in opposite directions, and fragmentations of a cell to multiple pieces. Mutagenesis studies have revealed that cells exhibit these modes depending on which genes are deficient, suggesting that seemingly different modes are the manifestation of a common mechanism to regulate cell motion. In this paper, we propose a hypothesis that the positive feedback mechanism working through the inhomogeneous distribution of regulatory proteins underlies this variety of cell locomotion and cytofission. In this hypothesis, a set of regulatory proteins, which we call cortical factors, suppress actin polymerization. These suppressing factors are diluted at the extending front and accumulated at the retracting rear of cell, which establishes a cellular polarity and enhances the cell motility, leading to the further accumulation of cortical factors at the rear. Stochastic simulation of cell movement shows that the positive feedback mechanism of cortical factors stabilizes or destabilizes modes of movement and determines the cell migration pattern. The model predicts that the pattern is selected by changing the rate of formation of the actin-filament network or the threshold to initiate the network formation.

**Funding:** This work was supported by grants for the 21st century COE program for Frontiers of Computational Science and by grants from the Ministry of Education, Culture, Sports, Science, and Technology, Japan.

**Competing Interests:** The authors have declared that no competing interests exist.

* E-mail: shin@tbp.cse.nagoya-u.ac.jp

## Introduction

Dynamical assembly and disassembly of intracellular actin filaments play important roles in the shape change of eukaryotic cells and in their locomotion [1]. For cell motility being stimulated by the external chemical signals, molecular mechanisms of regulatory dynamics of actin filaments have been intensively studied [2,3]. Even when there is no obvious external chemical signal, however, cells can move spontaneously in a randomly chosen direction [4]. Since the ability of spontaneous cell movement should be a basis for chemotactic responses, it is important to investigate the underlying mechanism. In this paper, we develop a theoretical model of spatio-temporal dynamics of actin filaments to reveal the mechanism of spontaneous behaviors.

In spontaneous movements, cells often take a "polypodal" shape by extending several pseudopods as can be found in a variety of cell types including a cellular slime mold *Dictyostelium descoideum* and macrophages in vertebrates. Their polypodal shapes are termed amoeboid because they resemble large water amoeba, *Amoeba proteus* [5]. Some other cells move spontaneously without taking the polypodal shape but by exhibiting a "crescent" shape. Fish epidermal keratocytes are examples of this type of cells [6]. *Dictyostelium discoideum* cells lacking *amiB* gene take the keratocyte-like shape [7], suggesting that amoeboid and keratocyte-like types are altered to each other by a minor change in biochemical reactions.

Variety of spontaneous movement is not limited to the above cases. In usual cytokinesis of animal cells, a contractile ring of actin and myosin II divides a cell into two daughter cells. *Dictyostelium discoideum* cells lacking myosin II, however, exhibit a cell-cycle-coupled division without a contractile ring through a process that two daughter cells crawl to opposite directions [8,9]. Cell division with the contractile ring is called "cytokinesis A" and cell division induced by the amoeboid crawling movement without the contractile ring is called "cytokinesis B" [8–10]. Furthermore, when the large, multi-nucleate cells are put on a substrate, they form multiple leading edges, which tear the cell into fragments in a manner uncoupled to the cell cycle [11]. Uyeda and his colleagues found that *Dictyostelium discoideum* cells lacking not only myosin II but either AmiA or coronin exhibit this type of cell-cycle-independent division, which was classified into "cytokinesis C" [10,12]. Since such cytofission is driven by the amoeboid crawling of cells, we may expect that the unified mechanism underlies both spontaneous cell locomotion and cytofission.

There are a lot of ways to treat large deformation of cell shape mathematically. An efficient way to reduce the computational cost is to consider only the boundary of a cell body. Stéphanou et al. [13,14] expressed a cell boundary by introducing a two-dimensional polar coordinate system, based on the two-phase model of Alt and his colleagues [15]. In this method the boundary of a two-dimensional cell was expressed by distance from a center point as a function of angle. Satulovsky et al. also used a similar polar coordinate expression based on the local-activator-global-inhibitor model [16], but the polar coordinate system cannot express shapes whose center is out of the boundary. Another way

**Author Summary**

Actin is a globular protein, assembling (polymerizing) into filaments. This process is called actin polymerization. Cell biologists have revealed that actin polymerization plays a central role in eukaryotic cell locomotion. Stimulated by internal/external molecular signals, actin polymerization occurs just beneath the cellular membrane. Such actin polymerization gives rise to pressure to push the cellular membrane outwards, which pulls the cell body and induces cell locomotion. Here, an important question on the mechanism is how the area of actin polymerization in cell is determined. To answer this question, we introduce a simple computational model that includes actin and a control factor of actin polymerization, which we call "cortical factor". Cell shape deformation induces heterogeneous distribution of cortical factor, leading to the heterogeneous actin polymerization in cell, which further enhances cell shape deformation. This feedback mechanism consistently explains a variety of modes of spontaneous cell movement, including both cell locomotion and cell division-like behaviors. Those different modes of movement emerge depending on the rate of actin polymerization and the threshold of concentration of cortical factor to control actin polymerization.

of expressing cell boundary is to use the level set method (LSM), in which the boundary of a cell is defined by a closed contour in a potential function [17]. Those methods to consider only the boundary, however, are not convenient to consider chemical reactions in cell body. In order to treat a whole cell body, one should consider elastic or fluid mechanics of the continuous media. There are two ways to describe the mechanics, Euler and Lagrange descriptions. With the Euler description, chemicals in a cell and the cell shape are observed at locations fixed in space, but with the Lagrange description, cell is tracked as a specific body. In many examples of modeling, the Lagrange description has been adopted by treating cell as a viscoelastic body. With the Lagrange description, Rubinstein et al. [18] constructed a two-dimensional model of fish epidermal keratocyte, with which the local density of actin and myosin within a cell was calculated to explain the displacement vectors of cell. Immersed boundary methods (IBM) is a variation of Lagrangian models with which the elastic bonds of actin filaments are treated together with the fluid dynamical description of cell medium [19]. In the many-particle model of Lenz [20], elastic bonds in membrane were also considered. Discrete models such as cellular automata, on the other hand, provide quite simple methods, which can largely reduce the computational cost. For example, Satyanarayana et al. developed a simple expression of cell shape, in which membrane was defined as a "chain" on the lattice space and actin proteins were treated as particles moving between lattice points [21]. In discrete models, a cell body can be defined by a set of connected lattice points, with which the use of Euler description is rather natural. For example, Marée et al. [22] explained keratocyte's locomotion by using a cellular Potts model (CPM) [23], in which the volume of a cell body was controlled by an energy-like cost function. Though those theoretical attempts explained important features of cell locomotion and deformation, unified treatment of both cell locomotion and cytofission has not yet been quantitatively discussed. In this paper, we develop a theoretical model to propose hypothesis that a single mechanism underlies a variety of different modes of movement, including amoeboid and keratocyte-type locomotion and cytokinesis B and C-type fission.

A unified description of cell locomotion and cytofission dates back to the review paper of Bray and White on cortical flow [24]: At the front edge of moving cell, actin is actively polymerized into the branched network and various protein factors such as Arp2/3 or uncapping proteins, which activate actin polymerization, are accumulated. Apart from the front edge, polymerization of actin network is somehow inhibited by accumulation of other protein factors, so that actin filaments remain to form skeletal structure at the cortical layer of cell [1]. As cell moves forward, this cortical actin is sent to the rear of cell in a manner similar to the flow of a caterpillar track and is dissolved into cytosol at the rear edge of cell (Figure 1a). Such a concerted flow of cortical actin has been called "cortical flow" and proteins which interact with actin filaments should be transported to the rear by this cortical flow. In the case of cytofission, the cortical flow runs from the front to the equator of cell and there cortical actin is dissolved into cytosol. Bray and White pointed out that cortical flow should play decisive roles not only in amoeboid locomotion but also in cytofission [24]. This cortical flow should give rise to the inhomogeneous distributions of bundled actin in the cortical layer and proteins that can interact with this cortical actin. When cell moves on a substratum, the bottom side of cell adheres to the substratum, so that the freely running cortical flow is absent on the bottom side. Even in such a case, cell movement should bring about the inhomogeneous distribution of cortical actin and other proteins as was suggested by Bray and White [24], and we here focus on such inhomogeneous distributions of proteins as a basis of unified description of locomotion and cytofission.

In previous papers, we have discussed the feedback mechanism which assures persistency in cell movement by developing a coarse-grained model of cell locomotion [25,26]. In this paper, we revise our model and treat both cell locomotion and cytofission within a unified framework by introducing the "cortical factor feedback model". We show that a variety of movements can be reproduced with this model through the feedback mechanism by changing the parameter to represent the speed of formation of actin filament network and the parameter that controls the spatial distribution of the network.

## Methods

### Cortical Factor Feedback Model

In this paper large deformation of cell is simulated to study both cell movement and chemical reactions on the same footing. In order to treat such large scale cell deformation, computational efficiency is an important requirement. Such efficiency is fulfilled by coarse-graining variables to be calculated. Since we need to coarse-grain dynamical rules among those variables, we do not consider here the detailed balance among mechanical forces explicitly but instead, we adopt the simplified kinetic rules of reactions and cell deformation.

Our coarse-grained description is based on the model of cell polarization. When cell is guided by the gradient of chemoattractant, cell is polarized upon receiving the chemoattractant molecules at the cell surface: Receptors at the cell surface initiate a cascade of events by stimulating the intracellular signaling molecules, which leads to a distinctive localization of signaling molecules in a polarized manner in a cell. These signaling events finally activate regulators such as Arp2/3 complex, which then stimulates the nucleation for actin polymerization. Growth of the actin filament network induces protrusion of the leading edge, which pulls the cell body forward. In this way, accumulated at the front side of cell are the branched actin network, Arp2/3 complex, proteins which uncap the barbed end of actin filaments, and other regulatory proteins to enhance polymerization of actin filament [27]. At the rear side of cell, on the other hand, actin filaments are bundled to form skeletal structures. Myosin II is accumulated at the rear and the actin-myosin complex generates the mechanical force to retract the rear of cell. Thus, at the rear side of cell, cortical actin, capped ends of actin, myosin II, and other regulatory proteins are accumulated and collectively work to inhibit formation of a branched actin network and the actin nucleation sites. This polarization stabilizes the directional motion of the cell to ascend the gradient of chemoattractant.

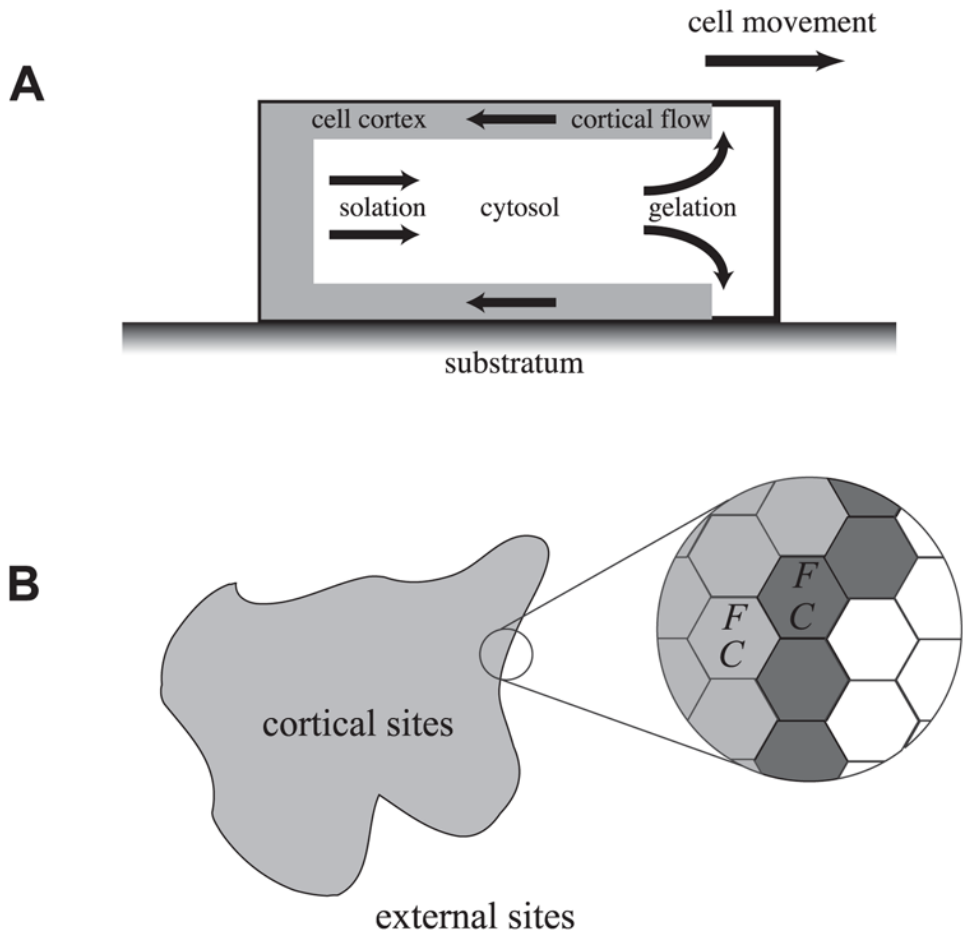

**Figure 1. The model of cell movement.** (a) A schematic view of dynamics of cortex layer. Cell cortex appears from cytosol by gelation through the formation of the branched actin network, flows into the rear edge of cell as cortical actin, and dissolves into cytosol by solation. Cell moves from left to right of the figure. (b) By neglecting the height of cell, cell movement in (a) are modeled by the two dimensional hexagonal grid. Cell body is represented by cortical and membrane sites in this grid. Right is a zoom-up view of the left picture. Gray and white hexagonal sites indicate cortical and external sites, respectively. Dark gray sites are membrane sites. Each cortical site has a set of two concentrations of cortical factor $C$ and actin filaments $F$.
doi:10.1371/journal.pcbi.1000310.g001

In the case there is no external chemical guidance, the spontaneous movement should be stabilized by a similar but spontaneously formed polarization of cells. In fact, many regulatory components for cell locomotion are localized spontaneously in a polarized manner along the length of a moving cell under no external cues [28]. To describe such stable polarization, we focus on protein factors which are accumulated at the rear and call them "cortical factors". Among cortical factors we include proteins that inhibit formation of branched actin network and interact with cortical actin, and cortical actin itself. When cell moves forward, these cortical factors are diluted at the front and accumulated at the rear of cell. In the present coarse-grained model, cortical factors are collectively represented by a single variable. Although more precise descriptions of multiple variables which are accumulated at the rear should improve the model, we use a variable of cortical factor to represent the feedback effects in an efficient way in the present model.

We assume a flat substratum and a flat cellular membrane by neglecting the height from the surface, which leads to the two-dimensional model of cell. Cell is modeled on the two-dimensional plane that consists of discrete hexagonal sites. A cell is defined by a set of connected sites in this space (Figure 1b). We call those sites "cortical sites", whereas other non-cellular sites are "external sites". Cortical sites which are adjacent to at least one of external sites are called "membrane sites". Cortical sites represent the side of cell that attaches to a substratum via adhesive molecules although we do not treat those molecules explicitly. In this model, the cell does not slide on the substratum but proceeds by creation of new adhesive bonds at the front and detachment at the rear of cell.

We assume that each cortical site can have two chemical species: Branched network of filamentous actin and the cortical factor, local concentrations of which are indicated by $F_j$ and $C_j$, respectively, where the suffix $j$ specifies the site position. We define the following rules:

**(1) Reaction kinetics.** The rule randomly selects a cortical site, $j$, and then updates $C_j$ and $F_j$ as follows:

$$C'_j = C_j + \beta - k_\beta C_j \qquad (1)$$

$$F'_j = F_j + \begin{cases} \gamma - k_f F_j & (\text{if } C_j < \alpha \text{ and } j \in \text{membrane}) \\ -k_f F_j & (\text{otherwise}) \end{cases}, \qquad (2)$$

where primed values in the left side of equations are the updated values. $\beta$ is the rate of transferring cortical factor from cytosol to cortical layer and $k_\beta$ is the rate constant of the reverse process. $\gamma$ is the rate of forming the actin network, $k_f$ is the rate constant of degradation of the actin network, and $\alpha$ is the threshold of actin polymerization. In Eq.2, the actin network is assumed to be formed only at the peripheral of cell *i.e.* at membrane sites.

Since Rho-associated proteins, which inhibit the actin-network formation, and Cdc42, which promotes the actin-network formation, are mutually inhibited [29–34] and the similar mutual inhibition can be expected between other proteins in cortical factor and the actin-network formation, it is reasonable to assume that promotion or suppression of actin-network formation is cooperatively dependent on the concentration of cortical factor. We thus can expect that the rate of actin-network formation at site $j$ is a sigmoidal function of $C_j$. In Eq.2, such a sigmoidal dependence is approximately treated by a step-functional on/off of the rate of actin-network formation, $\gamma$.

**(2) Diffusion.** When cortical factors bind loosely to the cortical layer, cortical factors should exhibit slow diffusion relative to the substratum. Here, we represent such slow diffusion by the following rule: The rule selects a cortical site, $j$, and then updates $C_j$ as

$$C'_j = C_j - nDC_j/6, \qquad (3)$$

$$C'_i = C_i + DC_j/6, \qquad (4)$$

where the $i$th site is a cortical site next to the $j$th site and $n$ is the number of cortical sites adjacent to the $j$th site. $D$ is a constant to determine the rate of diffusion. The rule executes Eq.4 for all $i$ around the $j$th site at one step. $D$ should be less than 1 by definition.

**(3) Cellular domain extension.** This rule simulates the observed mechanism that the increase in the amount of actin filaments leads to protrusion of the leading edge. First, the rule selects a membrane gird, and if $F_j$ in the selected $j$th site is larger than a certain threshold $F_{th}$, then an external site which is adjacent to the selected site is turned into a cortical site. Both the selected membrane site and the newly created cortical site share molecules by taking a half of the value of $F_j$ to represent conservation of mass of $F$. When there are more than one external sites adjacent to the selected membrane site, the rule randomly chooses one site from them.

Since cortical factor should have the smaller binding affinity to the branched actin network and should strongly bind only to the cortex that is fixed to the substratum by adhesion, we assume that the cortical factor is not pushed into the newly created cortical site with the extending actin filaments. Thus, whereas the mass of $F$ is split, $C_k$ in the newly created cortical site $k$ is set to zero.

**(4) Maintaining cellular body.** Cell shape dynamics should be determined by the balance among mechanical forces and chemical forces. Tensile forces in cortex and forces acting between cell and substratum are important mechanical forces and positive or negative pressures arising from the intra-cellular actin dynamics are chemical forces. In the present discretized model, however, it is not straightforward to describe the balance among forces in an explicit way. Instead, we here adopt the phenomenological rule by introducing a cost function.

The cost function is defined by $E = (V - V_0)^2 + cL^2$, where $V$ is the number of cortical sites, $V_0$ is the target cell size, $L$ is the number of membrane sites, and $c$ is a stiffness-like factor. First, the rule randomly selects a membrane site and randomly selects the operation of "adding" or "removing". If "adding" is selected, a new membrane site is created at one of the empty site adjacent to the selected site. $F$ and $C$ in the newly created site are transferred from a nearest neighbor cortical site. When there are multiple candidate sites from which $F$ and $C$ are transferred, one of them is selected randomly. If "removing" is selected, $F$ and $C$ of the selected site are transferred into a nearest neighbor site to satisfy the mass conservation, which leads to the increase of $F$ and $C$ there. When there are multiple candidate sites into which $F$ and $C$ are transferred, one of them is selected randomly. In this way, $F$ and $C$ are redistributed to reflect conservation of mass of them.

The above adding/removing operation is a trial operation and is accepted or rejected according to the Metropolis-like criterion: The trial is accepted with probability 1 when $E' \leq E$ and with the probability $P = \exp(-(E' - E)/T)$ when $E' > E$, where $E'$ denotes the cost function after the trial and $T$ is the parameter to determine the strength of fluctuation. If the removal of a site splits a cell into two or more disconnected domains, the execution is canceled and the other membrane site is chosen. The similar cost function was used by Marée et al. [22] to control the cell size in their model.

This rule is based on the assumption that the cell size tends to be kept constant during the cell movement. Such a global constraint on the whole cell size should be a natural consequence of approximately constant mass of cell and has been indeed observed in experiments of Karen et al. [35]. Karen et al. have shown that each motile epithelial keratocyte from fish does not change its total area during its motion. In this way, the term $(V - V_0)^2$ in the cost function is reasonable at least in the first order approximation. Resting cells, on the other hand, often exhibit rounded shapes because of their cortical tension [36]. If the cortex around a cell body is assumed to be simply elastic, contribution of the cortical tension to the energy should be proportional to $L^2$, which appears as the second term in our cost function $E$. By using this cost function, we represent effects of the mechanical forces. Then, the cell behaviors are determined by the balance between the constraint arising from $E$ and the protruding pressure of actin-network formation. The latter strongly depends on parameters $\alpha$ and $\gamma$ in Eq.2 and as described in the next section, diverse cell behaviors appear as $\alpha$ and $\gamma$ are altered. As explained in *Discussion*, such dependence of cell behaviors on $\alpha$ and $\gamma$ is not sensitive to the values of $c$ and $T$ in the present rule. This robustness of the model shows that the balance between mechanical and chemical forces is consistently described in the present phenomenological rule of maintaining cell body.

**(5) Sampling.** This rule has a role of clock for asynchronous updating procedures in the model. If this rule is called once, we count a simulation time step.

### Parameters

Parameters used in the model are summarized in Table 1. The time length of one step is assumed to be $\delta t = 1.0$ s. The length of a site is set to be $\delta x = 1$ μm, and the initial shape of cell is put to be a circle with 30-site diameter, corresponding to the typical size

**Table 1.** Parameter used in this paper.

| Parameter | Meaning | Values |
| --- | --- | --- |
| $\beta$ | Rate of transferring cortical factor from cytosol to cortical layer | 1 |
| $k_\beta$ | rate constant of transferring cortical factor from cortical layer to cytosol | 0.04 |
| $\alpha$ | Threshold of actin polymerization | 0.1–0.7 |
| $\gamma$ | Rate of forming the actin network | 1.5–4 |
| $k_f$ | Rate constant of degradation of the actin network | 0.99 |
| $D$ | Constant to determine the rate of diffusion | 0.45 |
| $F_{th}$ | Threshold for actin to create a new cortical site | 1 |
| $c$ | Stiffness-like factor | 2.0 |
| $T$ | Constant to control the extent of fluctuation | 20 |
| $V_0$ | Target cell size | 900 sites |
| $\delta t$ | One time step for simulation | 1 s |
| $\delta x$ | Length of a site | 1 μm |

doi:10.1371/journal.pcbi.1000310.t001

(several 10 μm) of a neutrophil. The equilibrium volume is set to be $V_0 = 900$. We use the normalized dimensionless representation for concentrations $F$ and $C$ by putting $F_{th} = 1$ and $\beta = 1$. Each of above five rules is called with the probability $P_i$ with $i = 1-5$. We give the rate $R_i$ for the $i$ th rule as $R_1 = r_1 V$, $R_2 = r_2 V$, $R_3 = r_3 L$, $R_4 = r_4 L$ and $R_5 = r_5$, and define $P_i$ by $P_i = R_i / \sum_{i=1}^{5} R_i$. Since the cortical factor binds or constitutes the cell cortex, its diffusion should be slower than cytosolic proteins. The effective diffusion constant of the cortical factor is $D^{\text{eff}} \equiv (D/6)(1.0 \ \mu m)^2 / \delta t \times (r_2/r_5)$. By setting $r_2 = 0.03$, $r_5 = 0.01$, and $D = 0.45$, we have $D^{\text{eff}} \approx 0.23 \ \mu m^2/s$, which is of about two orders smaller than the typical diffusion constant of cytosolic proteins. We set $k_f = 0.99$, so that $F_j$ is approximately zero when $j$ is not in the membrane, and $k_\beta = 0.04$ to keep the inhomogeneity of the distribution of the cortical factor. In the real time unit, $k_f^{\text{eff}} = 0.99/\delta t \times (r_1/r_5) = 0.14 s^{-1}$ and $k_\beta^{\text{eff}} = 0.04/\delta t \times (r_1/r_5) = 0.0056 s^{-1}$ representing the fast change in the distribution of the branched network of actin and slow transfer of cortical factor into cytosol, which assures the persistent spatial gradient of cortical factor across the cell. Other parameters are set to prevent the actin filament from spreading too broadly along the membrane and the cortical factor from uniform distribution; $r_1 = 0.0014$, $r_3 = 0.0071$, and $r_4 = 0.0143$.

Spatio-temporal dynamics of the actin network is controlled largely by the threshold of actin polymerization and the rate of actin polymerization, where the former affects the spatial spreading of the actin network and the latter determines the temporal scale of dynamics of the network. We investigate modes of cell movement and cell morphology by changing the threshold of actin polymerization, $\alpha$, and the rate constant of actin polymerization, $\gamma$.

### Cell Deformation Induces Spatial Gradient of Cortical Factor

Cortical factor is diluted at the front due to Rule (3) and is accumulated at the rear due to Rule (4), which amounts to the gradient of cortical factor from front to rear. Note that the accumulated actin network due to Rule (4) is disintegrated by following Eq.2 of Rule (2). Disintegration of the branched actin network takes place at every cortex site but formation of the actin network is limited at the membrane sites having small enough $C$, so that the accumulated $F$ at the rear due to Rule (4) is readily diluted and does not give a significant effect on the global distribution of $F$. We emphasize that the inhomogeneity of distribution of $C$ in the global cell scale generated by accumulation of $C$ at the rear and dilution of $C$ at the front is essential to describe the global cell shape and various modes of large scale motion as explained in the next section. If we omit Rules (3) and (4) and only consider Rules (1) and (2), density of cortical factor reaches equilibrium $C^* = \beta/k_\beta = 25.0$ at every site. As will be exemplified in Figures 2a and 3a, Rules (3) and (4) induce inhomogeneity in the distribution of $C$ to be $C < C^*$ at the front and $C > C^*$ at the rear of cell.

## Results

### Modes of Cell Locomotion

By varying $\alpha$ and $\gamma$, we found two characteristic types of stable locomotion. Figures 2a and 3a show corresponding two series of snapshots of distribution of the cortical factor in a cell and Figures 2b and 3b show two tracks of cell locomotion. See also Videos S1 and S2. We refer to the locomotion shown in Figure 2 as the amoeboid-like locomotion and the one in Figure 3 as the keratocyte-like locomotion. In both two types, concentration of the cortical factor is lower at around the front of moving cell and higher at around the rear. This inhomogeneity can be explained by the feedback mechanism which we call the *cortical feedback mechanism*: As cell starts to move in a direction, addition of cortical sites at the front dilutes the cortical factor and removal of cortical sites at the rear concentrates the cortical factor. Thus generated inhomogeneity of distribution of the cortical factor prevents the cell from moving backward and further stabilizes the forwarding motion. In this way, once cell starts to move in a direction, the cell tends to keep moving in that direction for a while through this positive feedback of motion and reaction.

Difference between two types of locomotion is the degree of fluctuation: The simulated amoeboid-like locomotion is much more fluctuating than the simulated keratocyte-like locomotion. As shown in Figure 2, shape of the amoeboid-like cell dramatically changes between the long polarized shape and the rounded shape. In contrast, as shown in Figure 3, the keratocyte-like cell keeps a laterally long shape. This difference in fluctuation is similar to the observed difference between the wildtype *Dictyostelium discoideum* cells and the keratocyte-like AmiB-null mutants [7].

In amoeboid-like locomotion, the threshold of actin polymerization, $\alpha$, is small but the rate of actin polymerization, $\gamma$, is large, which leads to the rapid actin polymerization in a localized region in the cell. Once the local region happens to have a large enough $F_j$, then that part protrudes to lead the cell body. The cortical factor is diluted at that protruding region and is concentrated at the opposite side of the cell (see 250 s in Figure 2a), which further enhances the protrusion at the front and contraction at the rear. In this way, the cell shape is elongated and the directed cell movement is stabilized through the cortical feedback. However, since diffusion of cortical factor is comparable with the speed of cell movement, the region where the cortical factor is diluted is not instantly filled by the diffusing cortical factor but is kept diluted behind the moving tip of cell after the movement lasts for a certain duration. Then, concentration of the cortical factor can be smaller than the threshold in this spread region and the actin network begins to be formed. Actin polymerization in this somewhat wide region promotes the protrusion around this area, which makes the cell shape round and the cell movement is slowed down. Then, the cortical factor is diluted at every protruding front, which further widens the region of small concentration of the cortical factor (325 s of Figure 2a). At this stage of the rounded cell, if some localized region happens to have large $F_j$ in its fluctuation, the cell begins to move in that direction, then the positive cortical feedback leads to the elongated shape again (at 375 s). In this way, coupled oscillations of cell shape, speed of movement, and the cortical factor distribution are inevitable in amoeboid-like locomotion as a consequence of the cortical feedback mechanism. In keratocyte-like locomotion, on the other hand, $\alpha$ is large and $\gamma$ is small. Then, actin is polymerized in a wide area with a moderate speed, which forms a stable laterally-long moving front of the cell. Through cell deformation, the cortical factor is diluted in this wide spread region and is accumulated in the rear side of the cell. This coupled pattern of motion and the cortical factor distribution is stable enough to keep the direction of cell movement through the cortical feedback mechanism.

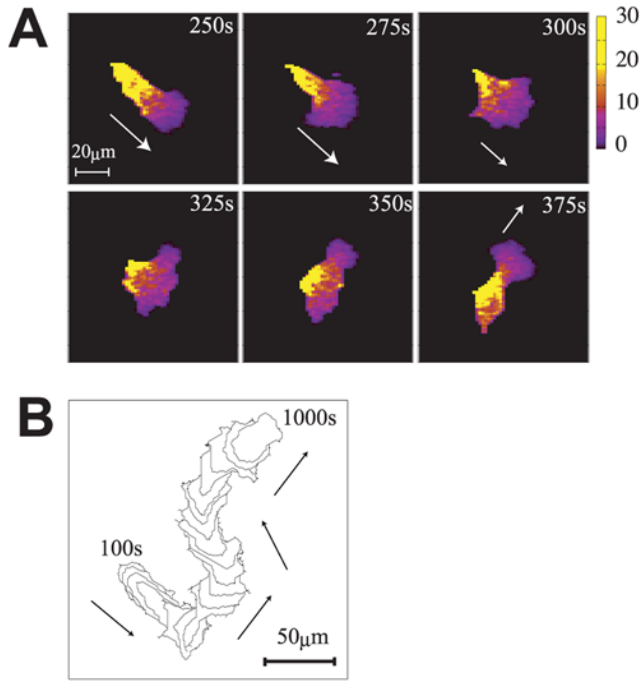

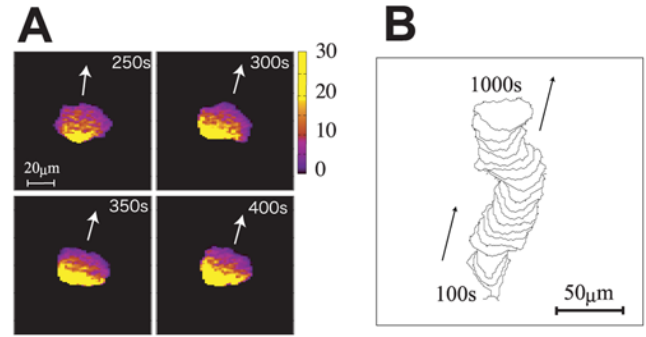

**Figure 2. Simulated amoeboid-like locomotion.** (a) Snapshots of the distribution of the cortical factor in the amoeboid-like locomotion. Parameters are set to $(\alpha,\gamma)=(1.8,3.8)$. Arrows in the panel indicate direction of motion of the cell. Colors indicate the concentration of the cortical factor. At the rear of the moving cell, the concentration of the cortical factor often exceeds its equilibrium value $C^*=25$. (b) A track of the amoeboid-like locomotion from 100 s to 1000 s drawn at every 50 s. The track at later steps masks the track of earlier steps. Arrows indicate the direction of motion.
doi:10.1371/journal.pcbi.1000310.g002

**Figure 3. Simulated keratocyte-like locomotion.** (a) Snapshots of the cortical factor distribution in the keratocyte-like locomotion. Parameters are set to $(\alpha,\gamma)=(7,1.6)$. Arrows in the panel indicate direction of motion of the cell. At the rear of the moving cell, the concentration of the cortical factor often exceeds its equilibrium value $C^*=25$. (b) A track of the keratocyte-like locomotion from 100 s to 1000 s drawn at every 100 s.
doi:10.1371/journal.pcbi.1000310.g003

## Statistical Analysis of Cell Locomotion

Differences between two types of locomotion can be quantified by measuring several statistical quantities. For example, the moving speed, $v(t)$, of center of mass of the cell should reflect oscillation of cell movement. Noisy high frequency component of $v(t)$ is filtered out when the moving average, defined by $\bar{v}(t)=\frac{1}{N}\sum_{t}^{t+N}v(t)$, is taken along the trajectory over $N=100$ s. $\bar{v}(t)$ shown in Figure 4a are the moving average taken along trajectories of Figures 2 (black line) and 3 (red line). We find the much larger fluctuation of $\bar{v}(t)$ in amoeboid-like locomotion than in keratocyte-like locomotion. In amoeboid-like locomotion, $\bar{v}(t)$ is larger when the shape is highly polarized at 250 s, and small when the shape is rounded at 325 s.

Inhomogeneity of the distribution of cortical factor in a cell is measured by $R_l(t)$, which is defined by the ratio of the number of cortical sites having $C_j$ lower than the average over the entire cell at the time step $t$. $R_l(t)$ is small when the cell is elongated and depletion of cortical factor is localized at the front edge, while $R_l(t)$ is large when the cell is rounded and cortical factor is diluted in a fairly large region of the expanding side of the cell. Figure 4b shows a scatter plot between $\bar{v}(t)$ and $R_l(t+100)$ in amoeboid-like locomotion, showing that both motion and reaction oscillate in a coupled way with the phase delay of about a hundred secs.

Directional persistence index $P_{dir}$ of cell movement can be measured by the average ratio of distance from a start point to the end point of motion of the center of mass of cell to the length of trajectory that the center of mass has traversed. The cell moves straight when $P_{dir}=1$ and the cell deviates from the straight path when $P_{dir}$ is small. In Figure 5a, $P_{dir}$ is shown in the $(\alpha,\gamma)$ space. When both $\alpha$ and $\gamma$ are small, cell is not strongly driven to move but is subject to fluctuations, leading to the random movement with less straightness. When both $\alpha$ and $\gamma$ are large, on the other hand, the random protrusion is amplified by the rapid actin polymerization and the cell tends to expand in a randomized way, which prevents the cell from showing the straight persistent movement. There is a domain of significantly straight movement from the left top to the right bottom of Figure 5a. $\alpha$ and $\gamma$ of the amoeboid-like and keratocyte-like cell lie at the left top and the right bottom of this domain, respectively, where the cell movement and chemical reactions are balanced to keep the straight movement. When we look more closely at this domain of relatively large $P_{dir}$, we find that $P_{dir}$ is larger in the right bottom than in the left top of this domain. In Figures 5c and 5d, we show that trajectories of the center mass of the cell are more straight in the the keratocyte-like locomotion than in the amoeboid-like locomotion. This straightness of the keratocyte-like locomotion can be confirmed in Figure 5a as the larger value of $P_{dir}$ in the parameter region of large $\alpha$ and small $\gamma$.

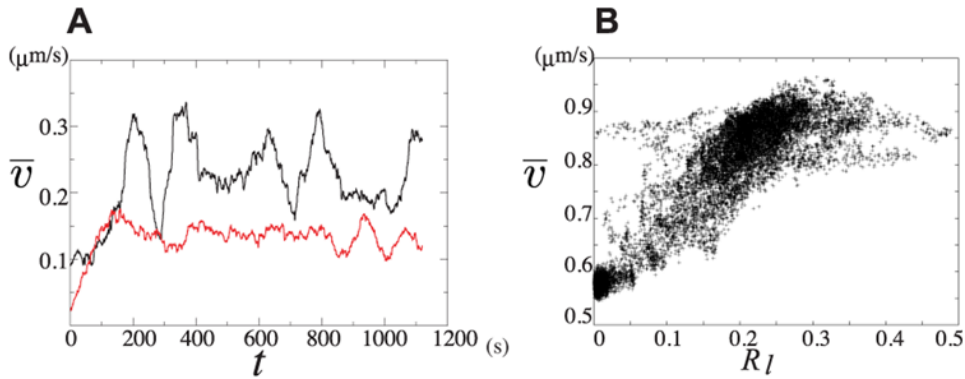

**Figure 4. Correlation between motion and reaction.** (a) Time series of the moving average of the speed of center of mass, $\bar{v}$, taken along the trajectory of the amoeboid-like motion of Figure 3 (black line) and $\bar{v}(t)$ taken along the trajectory of the keratocyte-like locomotion of Figure 4 (red line). $\bar{v}$ of the amoeboid-like type oscillates with a period of about 200 s. but $\bar{v}(t)$ of the keratocyte-like type does not oscillate significantly. (b) A scatter plot on the plane of $R_l(t+100)$, and $\bar{v}(t)$ of the amoeboid-like locomotion, where $R_l$ is the portion of area in which the concentration of cortical factor is low in a cell. The plot shows the strong correlation between $R_l(t+100)$ and $\bar{v}(t)$.
doi:10.1371/journal.pcbi.1000310.g004

Laterally long shape of the keratocyte-like cell can be detected by correlation $Corr$ between the direction of velocity and the direction of short axis of cell. $Corr$ is calculated by $Corr = \langle |\vec{m} \cdot \vec{v}|/|\vec{m}||\vec{v}| \rangle$, where brackets $<>$ indicates that average is taken both over 1000 steps interval in each simulation run and over 24 runs started with different random-number seeds. $\vec{v}$ is the velocity of the center of mass of the cell, and $\vec{m}$ lies along the minor axis of the cell calculated by fitting an ellipse to the the cell shape. If the value of $Corr$ is higher than $\frac{1}{2\pi}\int_0^{2\pi}|\cos\theta|d\theta = 2/\pi \sim 0.6366$, the cell tends to move along the minor axis. If $R \sim 0.6366$, there is no correlation between the minor axis and the velocity of the center of mass of the cell. (Note that zero does not mean no correlation.) Figure 5b shows $Corr$ as a function of $\alpha$ and $\gamma$, which indicates that the laterally long, keratocyte-like shape appears around the right bottom. Around the left top, $Corr$ is about 0.5, corresponding to the coexistence of two phases of the long polarized shape of $\cos\theta < 0.6366$ and the rounded shape of $\cos\theta \sim 0.6366$.

## Modes of Cytofission

As explained in the last section, rules of the model prohibit a cell from dividing into pieces. Nevertheless, the cell sometimes takes forms having distinct domains connected by narrow channels or cables. Although our model does not treat cell cycle, we found that these phenomena are morphologically similar to cell division. There are two types of cell division-like motion in the model. One is referred to as the cytokinesis B-like pattern and the other is referred to as the cytokinesis C-like pattern. See also Videos S3 and S4. In both two patterns, the cortical feedback mechanism plays important roles as explained below.

A time series of snapshots of the cortical factor distribution in the cytokinesis B-like pattern is shown in Figure 6, where $(\alpha,\gamma)$ is set to (3.8,2.5). This parameter set is at the intermediate between that of the amoeboid-like locomotion and that of the keratocyte-like locomotion. As in the keratocyte-like locomotion, a wide spreaded region on the front side of the cell has low concentration of the cortical factor. This region of the low cortical factor concentration is, however, not as stable as in the keratocyte-like locomotion. With a fluctuating distribution, the cortical factor happens to penetrate into the wide region of the low cortical factor concentration as shown with an arrow head in the panel (60 s in Figure 6). This penetration of the cortical factor destabilizes the directed motion of cell and two parts in the cell begin to move in opposite directions as crawling two daughter cells to show the cytokinesis B-like pattern. Once the two parts start to move in opposite directions, cell division is continued through the cortical feedback mechanism and a thin connecting cable is left between two parts (200 s).

Probability of occurrence of the cytokinesis B-like pattern is calculated by regarding the cell shape as having the cytokinesis B-like pattern if the number of distinct parts connected by a narrow cable in a cell is exactly two. The number of simulated trajectories showing the cytokinesis B-like pattern at least for some duration in their trajectories is counted and the probability is defined by its ratio to the number of all tested trajectories. The probability is shown in the parameter space of $(\alpha,\gamma)$ in Figure 8a. This probability is significantly high along the line from the left top to the right bottom in the panel, which largely overlaps with the region of straight movement shown in Figure 5a. The probability of occurrence of the cytokinesis B-like pattern is highest in the middle of this region at which the cell has both characteristics of the amoeboid-like movement and the keratocyte-like movement and can not stay in one of these two locomotive states to show the cell-division like instability.

The cytokinesis C-like pattern as shown in Figure 7 appears when $(\alpha,\gamma)$ is set to (4.5,4.5). What should be paid attention to is that an erosion indicated by an arrow head in Figure 7 is created at periphery of the cell, and the erosion grows larger to split the cell into multiple domains connected by narrow channels. This behavior is quite similar to the observed cytokinesis C [12,37]. In the model, enlargement of erosion is accelerated by accumulation of the cortical factor at the erosive front. Contraction at the erosive part concentrates the cortical factor there through the cortical feedback mechanism, that further promotes the erosion. As shown in Figure 8b, the probability of occurrence of the cytokinesis C-like pattern is high when both $\alpha$ and $\gamma$ are large.

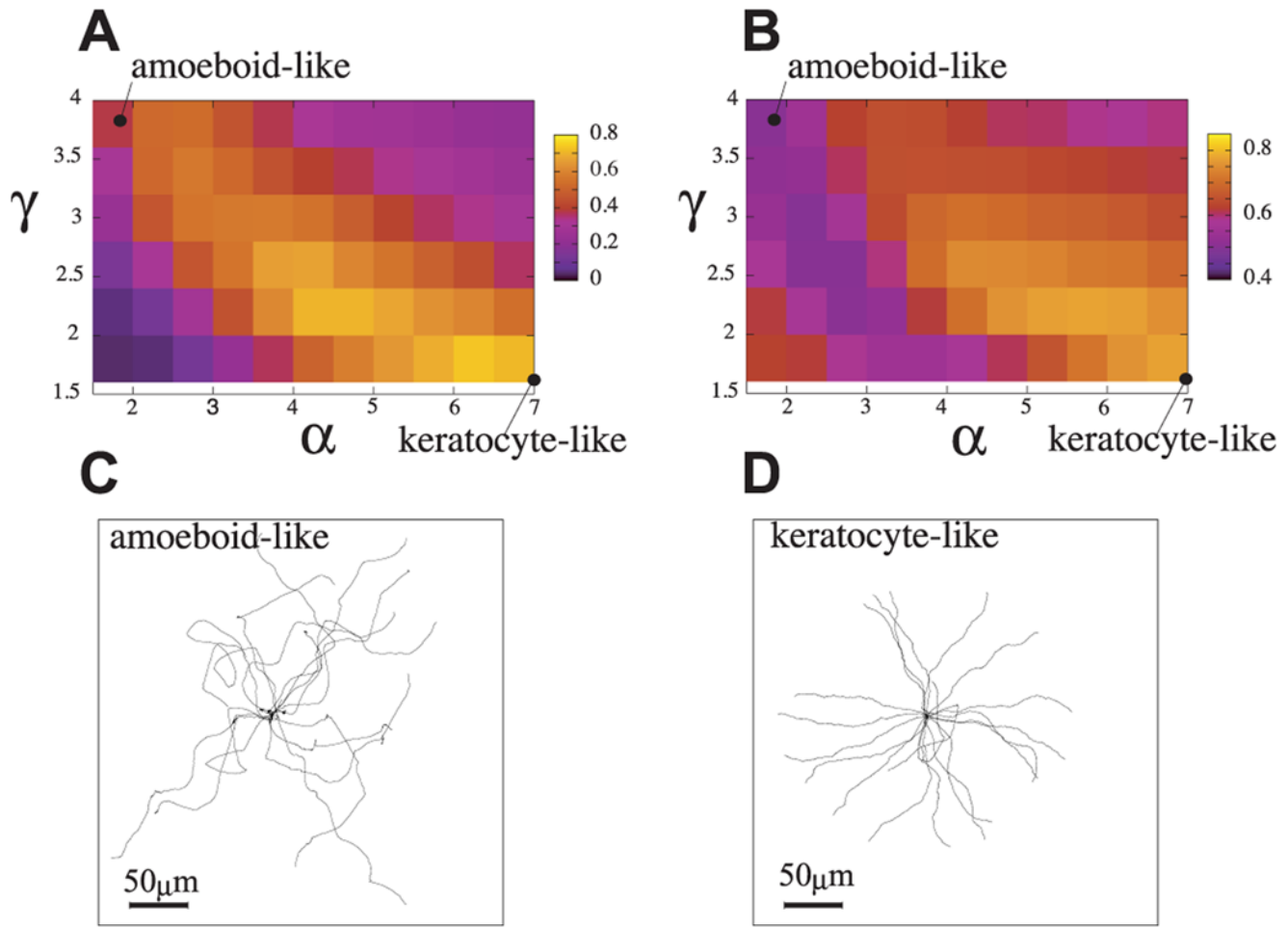

**Figure 5. Comparison between amoeboid and keratocyte-like locomotions.** (a) The color map of the directional persistence index, $P_{dir}$, on the plane of $\alpha$ and $\gamma$. $P_{dir}$ was measured by the average ratio of distance from a start point (0 s) to the end point (1000 s) of motion of the center of mass of cell to the length of trajectory that the center of mass has traversed. The value at each point in the color map is the average over 24 runs starting with different random-number seeds. The parameter sets for the amoeboid- and keratocyte-like locomotion, $(\alpha,\gamma) = (1.8, 3.8)$ and $(0.7, 1.6)$, are marked in the color maps. (b) The color map of the cell shape index, *Corr*, on the plane of $\alpha$ and $\gamma$. The average was taken over 1000 sec and 24 runs of different random-number seeds. See the equation in the main text for the definition of *Corr*. (c) Trajectories of the cellular center of mass of the amoeboid-like locomotion starting with different random-number seeds. Parameters are set to $(\alpha,\gamma) = (1.8, 3.8)$. (d) Trajectories of the cellular center of mass of the keratocyte-like locomotion. Parameters are set to $(\alpha,\gamma) = (7, 1.6)$.
doi:10.1371/journal.pcbi.1000310.g005

## Discussion

The cortical factor feedback model developed in this paper reproduced four typical patterns of movement. This ability of the model indicates that the cortical feedback mechanism, i.e. the motion-reaction feedback mechanism is the unified mechanism underlying a variety of patterns of spontaneous cell movement. This positive feedback stabilizes the straight movement in keratocyte-like locomotion, induces the oscillatory dynamics in amoeboid-like locomotion and destabilizes a single cell to split into multiple domains via cytokinesis B or C-like movement. Different modes of movements can be explained as variations in parameters that control the threshold and the rate constant of actin polymerization. Effects of modulation of the threshold should be experimentally tested by controlling the number of nucleation sites of actin polymerization in cell. Effects of modulation of the rate constant should be tested by regulating the concentration or affinity of proteins such as profilin, which binds to G actin to control the speed of actin polymerization.

Together with our model, biochemical and genetic evidence in regulatory mechanisms of actin polymerization may suggest a molecular basis for cortical factors. In the model, we referred to a collection of proteins which have an inhibitory role in actin polymerization as cortical factors. Then the model suggested that functional defects in cortical factors enhance formation of lateral pseudopods leading to destabilization of cellular polarity and motile persistence. In *Dictyostelium* cells, a series of mutant cell lines have been subjected to characterization of cell shape and motility [28]. A subset of mutants, including the null mutants of myosin II, clathrin, sphingosin-1-phosphate lyase, and PTEN, exhibit behavioral defects in which the mutant cells form a pseudopod more frequently from the lateral regions than the wild type and exhibit locomotion with less persistency, suggesting that these molecules are involved in the suppression of lateral pseudopods in a polarized cell. Some of them, *e.g.* myosin II and PTEN, may be cortical factors because those molecules are highly localized at the rear of a polarized *Dictyostelium* cell and around the equatorial regions of the dividing cell [38,39]. Accumulation of myosin II at the rear has been also reported in other cell types. Verkhovsky [40], for example, showed the accumulation of myosin II at the rear of moving fragments of a fish epidermal keratocyte cell. In Verkhovsky's experiment, cell movement was induced by the mechanical pushing at the initial moment, which strongly suggests that the accumulation of myosin II is not due to the chemical signaling but is induced by the cell shape deformation. The fact that myosin II acts as an actin depolymerization agent [41] also supports the idea that myosin II functions as a cortical factor. Since other regulatory proteins or cortical actin structure itself may also work as cortical factors, deletion of myosin II in mutants does not lead to the complete deletion of cortical factors but should alter the functionality of cortical factors, which can be reflected in the larger $\alpha$ in the model. Cytokinesis C-like movement explained by a large $\alpha$ in the model is consistent with the observed cytofission in myosin II-null *Dictyostelium discoideum*.

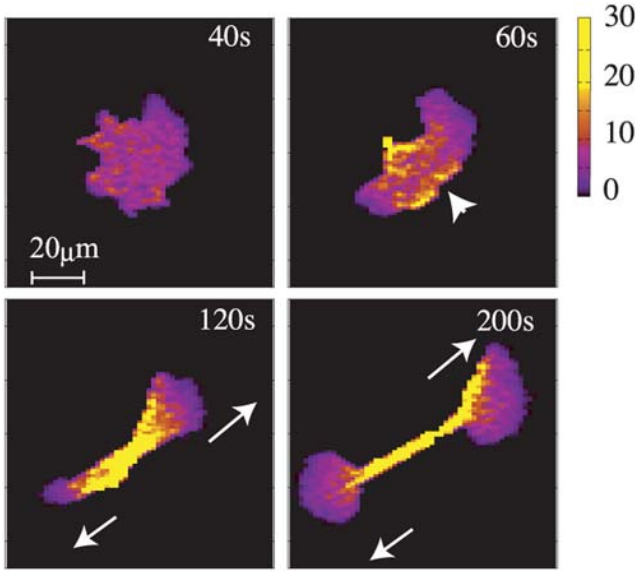

**Figure 6. Snapshots of the distribution of cortical factor in the cytokinesis B-like pattern.** Parameters are set to $(\alpha,\gamma)=(3.8,2.5)$. The arrow head in the panel of 60 s indicates the relatively high concentration of cortical factor at the cellular front.
doi:10.1371/journal.pcbi.1000310.g006

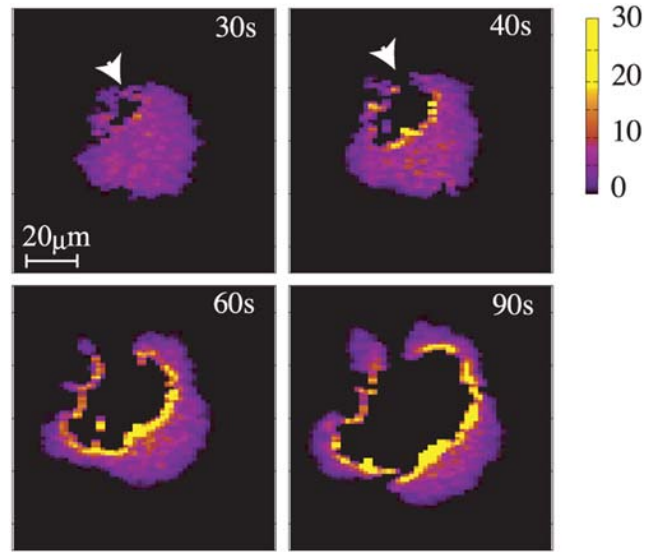

**Figure 7. Snapshots of the distribution of cortical factor in the cytokinesis C-like pattern.** Parameters are set to $(\alpha,\gamma)=(4.5,4.5)$. The arrow heads in panels of 30 s and 40 s indicate that an erosion appears and gradually grows.
doi:10.1371/journal.pcbi.1000310.g007

Another mechanism which can explain a variety of patterns of cell movement is the local-activator-global-inhibitor mechanism [16]. This mechanism may coexist with the cortical factor feedback mechanism of the present paper, but we should stress that cortical factors can dynamically change their distribution through change in cell shape or environment, so that the dynamical response of cell should be more appropriately explained by the cortical factor feedback mechanism. A similar mechanism of dynamical response was also discussed in the protocell model of Suzuki and Ikegami [42].

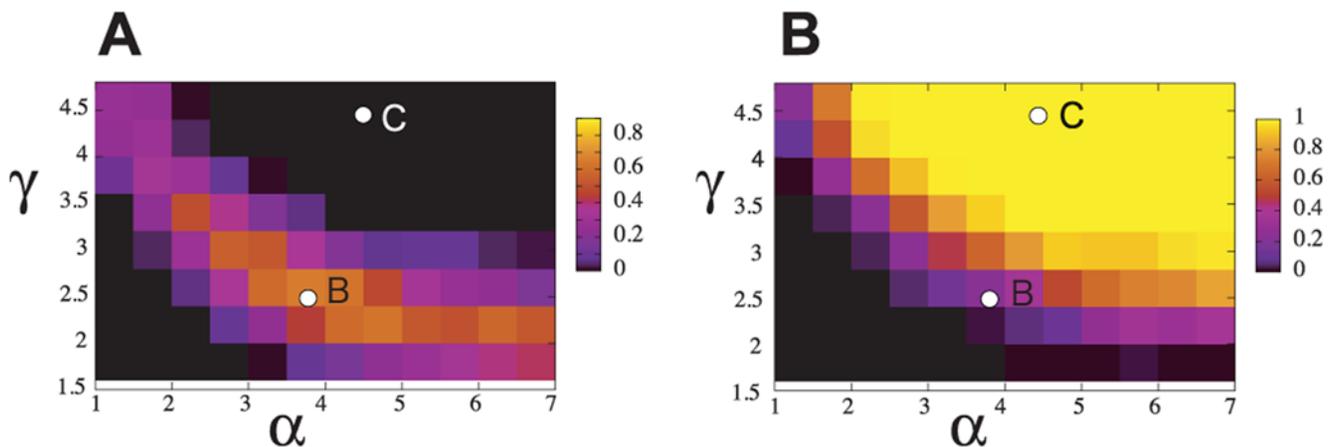

**Figure 8. Color maps of probabilities of occurrence of cytokinesis B- or C-like pattern.** The letters "B" and "C" in these maps indicate the corresponding parameters for cytokinesis B-like pattern of Figure 6 ($(\alpha,\gamma)=(3.8,2.5)$) and cytokinesis C-like pattern of Figure 7 ($(\alpha,\gamma)=(4.5,4.5)$), respectively. (a) The color indicates the probability of occurrence of the cytokinesis B-like pattern. (b) The color indicates the probability of occurrence of the cytokinesis C-like pattern.
doi:10.1371/journal.pcbi.1000310.g008

Cell shape dynamics should be determined by integrating balance of mechanical forces and chemical reactions at each local part of cell. In the present discretized model, integration of such local balance was not explicitly pursued but was replaced by many trials of updating sites under the Metropolis-like judgment. The cost function used in the judgment represents the constraint to keep the global cell size by making the peripheral length of cell small. A similar global constraint was successfully used in the model of Marée et al. [22] and the constraint was indeed observed in the experimental data [35]. Checking the robustness of simulated results against detailed changes of the constraint would further provide an evidence for the soundness of the constraint introduced in the model. We repeated simulations by changing $c$ and $T$ to examine this robustness. Increase in $c$ generates more rounded cell shapes in simulation, leading to the increase in the minimum value of $Corr$. The qualitative features of color maps of Figure 5 and 8, however, remain the same when $c$ is varied in the range of $1.0 \leq c \leq 4.0$. We also confirmed that color maps of Figure 5 and 8 are almost unchanged when $T$ is varied in the range of $10 \leq T \leq 80$, which showed robustness of the simulated results against changes in $c$ and $T$.

Extension of the present model to treat chemotaxis is an important next subject. Various modes of movement such as aggregation of *Dictyostelium discoideum* cells exhibiting an elongated shape were not treated in this paper but should be explained when the chemotaxis is taken into account in the model. In an immobile cell under the influence of external chemical cues, existence of the internal gradients of PI3K, PTEN, PIP3, and other proteins has been observed [43], which suggests that the intracellular chemical signaling works independently of whether the cell is moving or not. The cortical feedback, on the other hand, works through the cell movement. Interplay between the chemical signaling and the cortical feedback should further explain the complex behavior of cells induced by the external cues.

## Supporting Information

**Video S1** A video corresponding to Figure 2.
Found at: doi:10.1371/journal.pcbi.1000310.s001 (1.11 MB MOV)

**Video S2** A video corresponding to Figure 3.
Found at: doi:10.1371/journal.pcbi.1000310.s002 (1.10 MB MOV)

**Video S3** A video corresponding to Figure 6.
Found at: doi:10.1371/journal.pcbi.1000310.s003 (0.20 MB MOV)

**Video S4** A video corresponding to Figure 7.
Found at: doi:10.1371/journal.pcbi.1000310.s004 (0.12 MB MOV)

## Author Contributions

Conceived and designed the experiments: SIN. Performed the experiments: SIN. Analyzed the data: SIN MU MS. Contributed reagents/materials/analysis tools: SIN. Wrote the paper: SIN MU MS.